# TO THE CONSTRUCTION OF SYMMETRY-DEPENDENT INHOMOGENEITY LANDAU THEORY OF STATE AND ITS RELATIONSHIP TO FIELD THEORY

A.Ya. Braginsky

Research Institute of Physics, Rostov State University, Rostov-on-Don

e-mail: a.braginsky@mail.ru

**Abstract**

The paper shows that inhomogeneity of translation properties of condensate leads to the interaction of local order parameter (OP) and its compensating field which is similar to gauge fields in the field theory. Dynamic models were considered in the theory of phase transitions to inhomogeneous state with local translational symmetry, generalizing analogous models of static description [1, 2], where internal stress inevitably arises. In dynamic models, symmetry-dependent stress tensor defines force, and continuous group parameters of the local Landau potential are linked with the laws of conservation and equations of continuity. Equations of continuity for momentum in the present model, $\partial \sigma_{ij}/\partial X_j = \partial p_i/\partial T$, are in agreement with the second Newton's law and result from the equations of state, by analogy with equations of continuity for current, obtained from Maxwell equations in electrodynamics. It is shown that the principle of local homogeneity corresponds to the model of continuum with fibering of space into locally homogeneous subspaces in each of its points, resulting in the gauge model with minimal interaction in the Landau theory. Phenomenological charge responsible for minimal interaction of an order parameter (OP) and its compensating field linearly enters the expression for momentum, that being in agreement with the principle of equivalence of masses. Local Landau theory is consistent with ideas of quantum mechanics, because condensate, in a small volume, has wave properties and describes the wave function, which is transformed to local irreducible representation (IR) of the subgroup of translations.

## Introduction

A gauge model in the phenomenological theory of phase transitions was first introduced in the Ginzburg-Landau model which described the Meissner effect in the superconducting state [3]. Further application of a gauge model in the Landau inhomogeneous theory was suggested by de Gennes [4] to describe screening of the stress field in the deformed SmA, since the phase diagram for the nematic – SmA transition was equivalent to the phase diagram for transitions to the superconducting state for second-order superconductors. In both cases, the mechanism of interaction between the OP and its compensating field was borrowed from the field theory using the mathematical fact that a nonequilibrium potential allowed gauge transformation, i.e. simultaneous local transformation of phase of the complex OP and gradient transformation for the compensating field. As it is known [5], gauge models are characterized by non-trivial interaction which is written down in the form of an extended derivative (the so-called minimal interaction). Obviously, together with a compensating field, in an extended derivative additional curl gauge invariants emerged in the theory, composed of spatial derivatives of the compensating field itself. Thus, the nonequilibrium potential comprised three components: invariants of the matter fields (in the given case, of the OP or the wave function in the field theory); interactions of the matter fields and compensating fields in the form of extended derivatives, and invariants composed only of the components of the compensating fields, i.e. curl invariants. Whereas in the case of the superconducting OP, physical interpretation of curl invariants of the additional compensating fields was already known (namely, the magnetic field), to the contrary, in the case of Sm A, physical interpretation of the compensating field still had to be suggested [6, 7] (see Section 2).

On the other hand, given general formulation for an inhomogeneous model in the Landau theory of phase transitions, which was taking into account dependence of the OP $\eta^j = \eta^j(X)$ on the macrocoordinate and vector $k^l = k^l(X)$, continuous parameter of the IR, we obtained a model with the tensor compensating field [8]. The requirement of translational invariance of the local Landau potential resulted in emergence of the symmetry-dependent interaction of the local OP and the tensor compensating field in the model. Physical interpretation of the tensor compensating field of the OP allowed constructing the inhomogeneous Landau theory which necessarily takes into account the emerging stress field and density of dislocations in the inhomogeneous state [2].

To gain a better understanding of the nature of the obtained minimal interaction in the inhomogeneous Landau theory, in the present paper, we will consider the conservation laws which are directly associated with continuous symmetry groups of the nonequilibrium potential. To do that, we need to take into account time dependence of the OP and its compensating field in the model.

In the field theory [9], the law of conservation of charge is usually associated with the gauge symmetry group. It results from the equations of state obtained from variation of nonequilibrium potential with respect to the compensating field and takes the form of equations of continuity, such as, for example, charge conservation law in electrodynamics written down as an equation of continuity for current. In our case, the gauge group was induced by a subgroup of unit translations from the symmetry group of high-symmetry phase [2, 8]. At the same time, the subgroup of translations, as it is known [10], is in association with the momentum conservation law which can also be written down in the form of equations of continuity or Newton's equations. The fact that vector equations of continuity in the Landau gauge model (the model with tensor compensating field) are Newton's equations, is not obvious, but vector dimensionality of observable values and emergence of gauge structure stemming from the requirement of translational invariance point to it in an indirect manner. We will show (see Section 3) that the local gauge group of the nonequilibrium potential is connected with Newton equations of continuity which result from the subsystem of equations of state obtained from the variation of the nonequilibrium potential with respect to the tensor compensating field.

To integrate the notions such as gauge group for a potential and a subgroup of translations of the high-symmetry phase, we are required to reconsider conservation laws for the present model. In fact, in the inhomogeneous Landau theory, the integration of conservation laws exists, namely of conservation laws associated with unit translations, and laws of conservation of charges associated with the gauge group of nonequilibrium potential, and this is mathematically expressed in correspondence of Newton equations to equations of continuity for the gauge model. As no current can exist without field in electrodynamics, similarly in the present model, Newton equations can not exist without interaction, i.e. without minimal interaction between the OP and compensating field in an extended derivative. The new approach to a gauge group brings about new understanding of observables and coefficients in the field theory. Translational nature of the gauge group of the local Landau potential allows integrating the notions such as mass which is responsible for interaction, and inert mass which is proportional to the momentum (see Section 7). In other words, the law of conservation of momentum as of the first integral coincides with the law of conservation of charge which corresponds to the gauge group of nonequilibrium potential induced by unit translations, thus resulting in the principle of equivalence of masses in the studied model.

In constructing nonequilibrium potential in the inhomogeneous model, we will follow the Landau's principle, namely the requirement of invariance of the nonequilibrium potential with respect to the symmetry group of high-symmetry phase. The considered model is based on the principle of local homogeneity which is equivalent, as it will be shown (see Section 6), to the fibering of space into independent subspaces in each of its points, while the constructed gauge model is a result of the above fibering.

## I. Local homogeneity

The notion of local homogeneity was introduced in the Landau theory of phase transitions in describing phase transitions from homogeneous states to inhomogeneous ones where translational symmetry exists only in macroscopically small regions [2]. Inhomogeneous state which resulted from such phase transition is characterized by the fact that OP, in the spatially separated regions of the crystal, has different transformational properties. Indeed, if the difference in the distances between atoms in the inhomogeneous phase changes sufficiently slowly, then we can speak, with certain accuracy, about existence of a period in low-symmetry state in macroscopically small regions. Here, periods of unit cells in different macroscopic regions in the inhomogeneous state prove to be different. This is the physical mechanism for generation of dislocations and related stress at the junctions between locally homogeneous regions, the state of which is described by OP with local transformational properties [1].

Irrespectively to the theory of phase transitions, Kröner considered relaxation of stress in inhomogeneous medium through the onset of dislocations [11]. He pointed to a certain analogy between magnetostatics and continuum theory of stationary dislocations. However, in describing elastic properties of medium, the Kröner's approach did not disclose the origin of dislocations. In the present paper, we assume, as the driving force for onset of dislocations, relaxation of stress which appears in a crystal undergoing phase transition from the homogeneous state to the inhomogeneous one, with local translational symmetry on a macroscopic scale.

Usually, the transition to the inhomogeneous state is discribed by taking into account the dependence of the magnitude of the OP on the coordinates [12, 13]. We will focus on the case where not only the magnitude of the OP $\eta^l$ but also the vectors $k^l$ characterizing changes in the components of the OP $\eta^l$ under the action of translation operations on the period of lattice [1, 8] depend on the coordinates. Generally speaking, it is incorrect to write down the dependence of the OP on the coordinate, as OPs are coefficients for the corresponding harmonics in the Fourier series expansion of probability density $\rho(x)$. [14]. On the other hand, it is known that inhomogeneous phases do exist [12, 13]. To rule out of this contradiction, we need to formulate a local theory in the framework of the phenomenological Landau theory. It was suggested [8] to consider a model where the Landau potential can be defined locally, in a macroscopically small volume. It means that in a macroscopically small volume pointed out by macrocoordinate $X$, the local Landau potential $\Phi(X)$ exists, and it is the function of $\eta^l(X)$, $\partial \eta^l(X)$. Components of the OP $\eta^l(X)$ in each point of the continuum $X$ correspond to the coefficients in the expansion of the probability density into the Fourier series with respect to the microscopic coordinate $r$. To avoid confusion, we will hereinafter denote microscopic coordinate by $r$, and macrocoordinate by $X$. In this case we can speak about dependence of the components of the OP on macroscopic coordinates $X$ and formulate a variational theory in the macroscopic space $\{X\}$. It is worth noting that local homogeneity was used earlier in Refs. [3, 4, 12, 13], where the inhomogeneous Landau potential was also considered, with the only difference that $k=k(X)$ was not taken into account there.

It is evident that the spatial derivatives of the OP $\partial \eta^l(X)$ in the model with $k^l=k^l(X)$ are not eigenfunctions of the unit translation operator:

$$\hat{\vec{a}} \frac{\partial \eta_l}{\partial X_j} = \frac{\partial}{\partial X_j} \left( e^{i\vec{k}^l(\vec{x})\vec{a}} \eta_l(\vec{X}) \right) = e^{i\vec{k}^l(\vec{x})\vec{a}} \left[ i\frac{\partial \vec{k}^l(\vec{X})\vec{a}}{\partial X_j} \eta_l(\vec{X}) + \frac{\partial \eta_l(\vec{X})}{\partial X_j} \right], \quad (1)$$

while the OP itself is an eigenfunction of $\hat{\vec{a}}$ operator: $\hat{\vec{a}} \subset T_0$: $\hat{\vec{a}} \eta_l = \exp\{i k^l a\} \eta_l$. Here, $T_0$ is a subgroup of translations of the high-symmetry phase.

To construct the nonequilibrium Landau potential which describes the states with local translational symmetry and is invariant with respect to the subgroup of translations of the high-symmetry phase, it was necessary to introduce tensor compensating fields in the model [1, 8], which interact with the OP via minimal interaction [9], i.e. by introducing an extended derivative:

$$D_j^l \eta_l = \left( \frac{\partial}{\partial X_j} - i\gamma \sum_p A_{pj}^l \right) \eta_l, \quad (2)$$

which defines transformational properties of the compensating fields under the action of the translation operator:

$$\hat{\bar{a}}_q (\gamma A_{pj}^l) = \gamma A_{pj}^l + \delta_{pq} \frac{\partial k_p^l}{\partial X_j} a_q. \quad (3)$$

It is clear that this extended derivative (2) is the eigenfunction of the unit translation operator $\hat{\bar{a}}$. In Eqs. (2) and (3) $\gamma$ is a phenomenological charge; its physical meaning depends on the physical nature of the OP. Compensating fields introduced by Eqs. (2) - (3) allow consistently defining [1, 10] the stress tensor invariant to the operator $\hat{\bar{a}}$:

$$\sigma_{ij} = e_{jkn} (\partial A_{in} / \partial X_k). \quad (4)$$

The nonequilibrium Landau potential contains only antisymmetric linear combinations of derivative components of the compensating field $A_{ij}$ (4). Indeed, according to (3), antisymmetric combinations of spatial derivatives of the compensating field with respect to the second index, (4), are invariant with respect to the transformations from the subgroup of translations: $\hat{\bar{a}} \subset T_0$. The fields themselves enter $\Phi(X)$ only in form of real quadratic combinations of OP components invariant with respect to the unit translations, and their extended derivatives (2). Hence, the local Landau potential invariant with respect to unit translations depends on three types of invariants:

$$\Phi(X) = \Phi\left( \eta_l \eta_l^*, D_j^l \eta_l D_j^{*l} \eta_l^*, \sigma_{pi} \right). \quad (5)$$

## 2. Gauge transformation in the Landau theory

The first attempt to construct a model with $k=k(X)$ was made by de Gennes in Ref. [4], while describing the deformed SmA in the vicinity of the second-order phase transition nematic-SmA. He suggested inhomogeneous distribution of the director $n=n(X)$ in the low-symmetry phase as a result of external deformations. Since the director $n$ in SmA relates to the vector $k$ of the smectic OP by the relation

$$k=n/d, \quad (6)$$

where $d$ is the distance between the layers, then for $n=n(X)$ it is clear that $k=k(X)$.
De Gennes attempted to describe the effect of screening of the stress field in terms of dislocations occurring in the deformed SmA, by analogy with the Meissner effect for second-order superconductors [3]. However, he defined the elastic energy for the deformed SmA as the Frank potential:

$$\Phi_F = \frac{1}{2} k_1 (di\ \mathbf{n})^2 + \frac{1}{2} k_2 (\mathbf{n} rot \mathbf{n})^2 + \frac{1}{2} k_3 (\mathbf{n} \times rot \mathbf{n})^2 \quad (7)$$

which is known [15] as characterizing elastic properties of nematics rather than of smectics. The potential (7) for nematics does not take periodic structure of smectics into account (see Ref. [15]). In Ref. [4], de Gennes did not take into account the locality of transformational properties of the smectic order parameter with respect to the translation operator with the condition $n=n(X)$, which results in (1) according to (6). Thus, locality of the transformational properties of the OP in the de Gennes' model implies that the potential [4] is non-invariant with respect to translation operators, this being in contradiction with the fundamental principles of the Landau theory. If we employ a formalism which uses compensating fields, to construct the smectic Landau potential, then to solve the de Gennes' problem [6, 7], associated with the fact that the Frank potential contains non-curl summands (7), it will be sufficient to express the elastic energy for SmA in terms of the stress tensor (4), (5) as antisymmetric combinations of spatial derivatives of the compensating tensor field [16].

Consequently, the necessity to include compensating fields in the Landau theory of phase transitions arose due to the fact that the derivatives of the OP are not eigenfunctions of the translation operator (1) for any inhomogeneously deformed state with $k=k(X)$. Thus, to construct invariants of the nonequilibrium potential depending on the spatial derivatives of the OP, for example, $D_j^l \eta_l D_j^{*l} \eta_l^*$, it was required to introduce tensor rather than vector [4,6,7] compensating fields (2). Nonequilibrium potential as a function of the compensating fields must also contain their spatial derivatives. Invariants composed of the spatial derivative components of the compensating fields take the form (4). Since the vectors $k^l$ in the star of the IR are connected by the transformations of symmetry of the point group, then, similarly, the components of the compensating fields must also relate to each other. It was shown in Ref. [1] that it is sufficient to introduce a single compensating tensor field for all components of the OP and interpret it as the potential of the stress field (4) [1]. Equilibrium values of the OP and compensating fields are defined by solving the following system of equations of state:

$$\delta\Phi/\delta\eta_l = 0, \quad \delta\Phi/\delta A_{pj} = 0. \tag{8}$$

Equations (8) completely define the equilibrium state of the inhomogeneously ordered system. Equations of state $\delta\Phi/\delta A_{pj}=0$ in (8) are basic equations of the continuum theory of dislocations: $\rho_{ij} = e_{jkn}(\partial w_{in}/\partial X_k)$, where $\rho_{ij}$ is the density of dislocations, and $w_{in}$ is the tensor of elastic distortion [1,11,15] (see Eq. (14) below). The expression for the density of the Landau potential describing transition to the inhomogeneous state at $k=k(X)$ is similar to that for the Ginzburg-Landau potential in the theory of superconductivity. The difference lies in the tensor dimensionality of the compensating fields. Compensating fields in (8) have an additional vector index associated with the vector dimensionality $k$. The OP, generally speaking, is multicomponent and corresponds to the vector star $\{k\}$.

We did not use the notion of the local gauge group in our studies. On the contrary, we are speaking not about a local group but of a local representation, and introduce compensating fields as a necessary element in diagonalization of the basis for representation of spatial derivatives of OP; it enables to construct invariants for the subgroup of translations of the high-symmetry phase of a crystal. Thus, dependence both of the magnitude of the OP and its transformational properties at $k=k(X)$ on macroscopic coordinates results in the gauge model with minimal interaction between the OP and stress tensor potential, defined in terms of extended derivatives. In the present model a group is not local, but non-additive continuous parameters determining its representation, namely vectors $k(X)$ are local. This very feature allows to construct, in the continuum approximation, the gauge model even for the discrete Abelian group $T_0$. Mathematically, this can be expressed as:

$$\alpha(X) = k(X)a, \qquad (9)$$

where $\alpha(X)$ is a continuous parameter of the gauge group [5, 9]. Actually, here we are considering a way to derive an expression for the phase of gauge transformation (9).

We have shown in [1, 10] that the principle of locality for the OP with respect to the temporal translations $\hat{\tau}$, when the gauge group parameter takes the form:

$$\alpha(X) = \omega(X)\tau, \qquad (10)$$

would result exactly in the Ginzburg-Landau potential, and Maxwell equations in the system of equations of state. In this case, the vector potential transforms as follows:

$$\hat{\tau}(\gamma A_j) = \gamma A_j + \frac{\partial \omega}{\partial X_j}\tau, \qquad (11)$$

and changes its sign with the inversion of time (since $\omega(X)$ is the local parameter of representation of temporal translations). The change of sign of the vector potential of electromagnetic field upon inversion of time to postulate in the field theory [9], since there was no explicit expression for gauge transformation (10). In the inhomogeneous Landau theory, the change of sign of the vector potential upon inversion of time stems from the locality of transformational properties of the superconducting OP with respect to temporal translations.

## 3. Dynamic model and definition of momentum

Up to now we were describing static models somehow associated with physics of phase transitions. Incorporation of equations of elasticity and electromagnetics in the local Landau theory suggests interrelations between the present model and the field theory [3, 5, 9, 17]. Dependence of field functions and observed parameters on time plays significant role in the field theory. To construct a complete compensating 4-tensor, we need to consider the case of dependence of the parameters of the IR on time (here, the electromagnetic potential may formally be viewed as the fourth component of the tensor compensating field). In this case, the first indices of the compensating 4-tensor will correspond to spatial and temporal translation.

As it is known, the conservation law of momentum is linked, in classical mechanics, to translation operators [10]. It is evident that in the dynamic case we must obtain a consistent model with the conservation law related to the first integrals of the equations of state. Conservation laws can be written locally as equations of continuity. According to the E.Noether's theorem [9], continuous symmetry groups of the nonequilibrium potential are associated with equations of continuity. They include groups of transformations of space-time and so-called continuous groups of internal symmetries which are allowed by the constructed nonequilibrium potential. Local groups of internal symmetries are known as gauge groups. Invariance of nonequilibrium potentials with respect to them results in occurrence of minimal interaction between wave function and compensating field in the form of an extended derivative.

Groups of internal symmetries [5], which are associated with charge conservation laws, are, by definition, not associated with the space-time symmetry; rather, they act only in the space of functions. However, in the obtained local Landau theory, the unit translation operator $\hat{\bar{a}}$ acting at the physical point with the macrocoordinate $X$ induces the gauge group of the nonequilibrium potential (9) [1]. The question arises how to define the momentum in the local Landau theory?

As shown above, stress tensor necessarily appears in the inhomogeneous Landau theory. In dynamic models, it will determine force since divergence of the stress tensor is, by definition, a force. Let us show that equations of continuity analogous to equations of continuity in electrodynamics, take the form of the second Newton's law, and hence, the momentum conservation law, are connected with invariance of nonequilibrium potential with respect to the unit translation operator $\hat{\bar{a}}$.

(**Footnote:** Note that in non-Abelian models [5] it would be impossible to define force in this manner, since divergence of observed fields in the Young–Mills theory is, by definition, not a vector.)

Let us introduce the notion of macroscopic and microscopic time. For the microscopic time, we can suggest that characteristics of the OP (specifically, of the vector $k^l$ in the given point of $X$) remain constant. Let us assume that the dependence of the OP on macroscopic time is such that the equilibrium values of the observed parameters can be attained. In this case, a corresponding variational formalism can be constructed.

To define the stress tensor $\sigma_{ij}$ in [1] we used the expression

$$\sigma_{ij} = e_{jkn}(\partial \Sigma_{in}/\partial X_k), \qquad (12)$$

where the compensating field of the OP $A_{ij} \equiv \Sigma_{ij}$ is a curl potential of the stress field. Components of the stress tensor $\sigma_{ij}$ in the theory of elasticity are defined as values conjugated to components of the tensor of elastic distortion $w_{ij}(X)$ [15]:

$$\partial \widetilde{\Phi}/\partial w_{ij} = \sigma_{ij}. \qquad (13)$$

Here the potential $\widetilde{\Phi}(X)$ depends on the tensor of elastic distortion and it is conjugated to the potential (5).

Taking into account that translation-invariant density of the potential has the form $\Phi(X) = \Phi\left(\eta_l \eta_l^*, D_j^l \eta_l D_j^{*l} \eta_l^*, \sigma_{pi}\right)$, the equations of state $\delta\Phi/\delta A_{pj}=0$ define the density of dislocations:

$$\rho_{ij} = e_{jkn}(\partial w_{in}/\partial X_k), \qquad (14)$$

where

$$\rho_{ij} = \partial \Phi/\partial A_{ij} \qquad (15)$$

is the density of dislocations [15]. It corresponds to the first integral induced by the operator $\hat{\bar{a}}_i$. The first integral (15) is the linear combination of equations $H_j^l = \frac{\partial \Phi}{\partial(\partial \eta_l/\partial x_j)} \eta_l$, $H_j^{l*} = \frac{\partial \Phi}{\partial(\partial \eta_l^*/\partial x_j)} \eta_l^*$, multiplied by the phenomenological charge, according to Eqs. (2):

$$\rho_{ij} = -i\gamma \left[ h_i^l H_j^l - h_i^l H_j^{l*} \right]. \tag{16}$$

For example, this is the form of the density of dislocations constructed for quasicrystals with nontrivial icosahedral vector star $\mathbf{k}$ in Ref.[18]. Coefficients $h_i^l$ display linear relation between the vectors in the star $\{\mathbf{k}^l\}$ determined by their transformational properties under rotation operations from the symmetry class $G_0$, i.e. symmetry group of the high-symmetry phase. Due to this, all arms in the star of the compensating fields $\{A_{pj}^l\}$ are linear combinations of components of the single tensor compensating field $A_{pj}^l$ [1]. Hence, the density of dislocations (16) transforms as a vector with respect to the first index under transformations from the symmetry class $G_0$.

In order to construct a model taking dynamics of defects into account, let us consider the dependence of the vector $\mathbf{k}=\mathbf{k}(T)$ on the macroscopic time. In this case, it would be more correct to speak about construction of a Lagrangian. We retain the same terminology and speak about construction of the nonequilibrium potential which depends both on macroscopic coordinates and on macroscopic time. In the dynamic model, to define the stress tensor, besides the condition (13), we must also take into account that $\sigma_{ij}$ is a tensor such that its divergence is, by definition, a force [15]:

$$\partial \sigma_{ij}/\partial X_j = f_i \tag{17}$$

Thus, in the above model, we need to additionally define the stress tensor, as the divergence of that tensor defined according to (4) is zero. This can be done by introducing additional potential vector compensating fields analogous to the potential of the electric field in electrodynamics, and by taking into account temporal dependence in the tensor compensating fields.

Let us proceed by analogy with electrodynamics [9, 17] and extend the temporal derivative of the OP using the vector potential $\phi_i$:

$$D_0^l \eta_l = \left( \frac{\partial}{\partial T} - i\gamma \sum_i \phi_i^l \right) \eta_l \tag{18}$$

with the transformation properties:

$$\hat{\bar{a}}_q(\gamma \phi_i^l) = \gamma \phi_i^l + \delta_{qi} \frac{\partial k_i^l}{\partial T} a_q, \tag{19}$$

which, unlike the scalar electric potential, $\phi$, changes its sign upon the inversion of time. As a result of variation of $\Phi$ with respect to $\phi_i$: $\delta\Phi/\delta\phi_i=0$ we obtain potential equations in the system of equations of state in the following form:

$$\partial D_{ij}/\partial X_j = p_i, \tag{20}$$

where $p_i = \partial \Phi / \partial \phi_i$ is a vector changing its sign upon the inversion of time, and $D_{ij} = \partial \Phi / \partial E_{ij}$ is a value conjugate to the translationally invariant combination $E_{ij} = \partial A_{ij} / \partial T - \partial \phi_i / \partial X_j$ according to (3), (19) (we delete index $l$ here, since it was shown in [1,2] that it would be sufficient to compensate one arm of the star of the IR).

To obtain explicit dependence of the stress tensor on the potential fields, let us use dualism which exists in the stationary model [1, 11]. It lies in the fact that our theory contains two curl equations for the observed fields. One is the invariant of the subgroup of translations composed of the derivatives of the compensating field, while the second one is the equation of state obtained from the variation of the nonequilibrium potential with respect to that field, $\delta \Phi / \delta A_{pj} = 0$. If we define $A_{ij} \equiv \Sigma_{ij}$ as the potential of the stress field, then these will be the equations (12) and (14). However, we may equally define the compensating field as $A_{ij} \equiv w_{ij}$, the tensor of elastic distortion [1]. Then equations in (14) are invariants of the subgroup of translations of the potential, and (12) are the equations of state $\delta \widetilde{\Phi} / \delta A_{pj} = 0$ according to (13). Now, by shifting to the conjugate variables, we will show that the choice of elastic distortion $A_{ij} \equiv w_{ij}$ as the compensating field and the variable yields the model with the equations of continuity in form of the second Newton's law.

Indeed, in this case translationally invariant density of the nonequilibrium potential, according to (14) has the form $\widetilde{\Phi}(\vec{X}, T) = \widetilde{\Phi}(\eta_l \eta_l^*, D_j^l \eta_l^* D_j^l \eta_l^*, \rho_{ij}, E_{ij})$, and the first integral corresponding to the continuous parameters $\mu_i$, those are coordinates of the vector $\boldsymbol{k}$ in the reciprocal space, is defined, according to (13), similarly to the current in electrodynamics, and appears in the equations of state $\delta \widetilde{\Phi} / \delta A_{ij} = \delta \widetilde{\Phi} / \delta w_{ij} = 0$ as follows:

$$\sigma_{ij} = e_{jkn}(\partial \Sigma_{in} / \partial X_k) - \partial D_{ij} / \partial T. \tag{21}$$

Here $\partial \widetilde{\Phi} / \partial \rho_{ij} = \Sigma_{ij}$ is the curl potential of the stress field, $\rho_{ij} = e_{jkn}(\partial w_{in} / \partial X_k)$ is the density of dislocations defined by Eq. (14), and $D_{ij}$ is the value conjugated to the combination $E_{ij} = \partial A_{ij} / \partial T - \partial \phi_i / \partial X_j$. Applying the divergence operator to the equation (21) and differentiating (20) with respect to time, one obtains the equations of continuity in the form:

$$\partial \sigma_{ij} / \partial X_j = \partial p_i / \partial T. \tag{22}$$

The equations (22) allow only one interpretation of $p_i$ in Eq. (20), namely as the momentum, as its left-hand part in (22) is a force (17). It follows herefrom that the equations of continuity (22) represent the second Newton's law for the locally inhomogeneous dynamical model which describes the dynamics of defects in the lattice. I.e., in the local Landau theory, the fundamental law of dynamics is determined by invariance of the nonequilibrium potential with respect to the spatial translations onto the periods of the unit cell. It is worth noting that if we define $A_{ij} \equiv \Sigma_{ij}$, the Eq. (22) is also valid, and in this case it is required to define the potential part of the stress tensor connected to the field $\phi_i$, while in the equations of continuity obtained from the equations of state, conjugate values should appear.

We should also notice that the momentum (20) defined as a value conjugate to the vector potential, $\phi_i$, in the approximation of continuum, is actually a current. However, it is not a curl current [3], but a current associated with potential forces (20). Let us use formal analogy with electrostatics. As it is known, a conductor screens an external field. In our case, where the momentum $p_i$ is a function of state, it is a source of the internal field, $D_{ij}$, according to (20), which must screen the external field $E_{ij}$ in order to have a constant potential $\phi_i$ inside the conductor. It is clear that such a state cannot be sustained for a long time because the momentum is the current in the medium, and divergence of this current is not zero in the stationary case, according to (20). Thus, we should expect that the source of the above current will be exhausting, and screening of this kind cannot be long-standing. The state where electrically charged particles are the carriers of the momentum determined by the potential field, can also screen the external magnetic field. In the stationary case, the standard equation of continuity for charged particles: $\partial p_i/\partial X_i = \partial j_i/\partial X_i = \partial \rho/\partial T$ results in linear dependence of their density $\rho = \rho_0 - \partial p_i/\partial X_i \cdot T$ on time. This model is interesting in that the implementation of the process of screening of the external field limited in time and takes place in the state with local translation symmetry in which short-range order, rather than long-range order, exists. Such model is in good agreement with models of high-temperature superconductivity (HTSC).

## 4. Observables: defects and deformations

Let us point out the fact that in the model with local translation symmetry observables are defects, i.e. disjoints in the lattice [18]. This is a common property: observable values in the symmetry model appear as breaks of the symmetry. Let us consider distortion of a lattice as a special case of deformation of periodic structure. The simplest and the most illustrative example is a model of a smectic. In the case of a smectic, deformation of vectors $k^l(X)$ can be presented as deformation of the basis of the reciprocal space. In Ref. [16], distortion of the basis of the reciprocal space was actually considered, when the director deviated from the main optical axis: $n=n(X)$. With local rotation of the main optical axis, which takes place in the de Gennes' model [4], the density of defects $\rho_{ij}$ depends only on distortion $\delta n(X)$, and the surface integral $\rho_{ij}$ does not coincide with the Burgers vector [16]. Physical interpretation of this linear defect suggests that it characterizes the magnitude of misfit between the layers at $d=const$, since for small distortions, the Burgers vector is perpendicular to the variation $\delta n(X)$ [4]. Thus, it would be more correct to denote the tensor $\rho_{ij}$ in Eq. (14) as the tensor of elastic linear defects.

Another special case was considered in Ref. [18], where the basis was not rotating, whereas only the scale was varying: $d=d(X)$. We have already used such model to describe a deformed quasicrystal [18] where, instead of varying vectors $k^l(X)$, the magnitude of basis vectors of the reciprocal space was a variable parameter. It is interesting to note that in this case, the surface integral of the density of dislocations corresponds to the Burgers vector.

In the above examples, the continuous parameter of the IR, $k=k(X)$, was actually changing, whereas deformation of the basis of the reciprocal space was merely an interpretation. In the Landau theory, non-equilibrium potential must be invariant with respect to the symmetry of the high-symmetry phase in the homogeneous state. Dependence of the OP and its transformational properties on $X$ does not mean curvature of space itself, since, in the absence of the OP, the symmetry of state corresponds to the high-symmetry homogeneous phase of a crystal. From here we may conclude that in the inhomogeneous Landau theory, the principle of locality of continuous parameters of the IR, which can be viewed as deformation of the basis of the reciprocal space in the low-symmetry state, results in the gauge theory with minimal interaction (1), (2) [1, 2], but not in the formalism of the general theory of relativity (GRT) [17], as might be expected.

If we consider a lattice in form of three independent smectic models, where the director in Eq. (6) acts as the basis of the reciprocal space, then we obtain the picture for deformation of a 3-dimensional lattice:

$$e^i(X) = g^{ij}(X) e^j. \tag{23}$$

If the dependence $k^l = \mu_i^l e^i(X)$ on $X$ is given in terms of the tensor (23) then $g^{ij}(X)$ will act as the local parameter of representation in (9). In the given model, to construct the invariant non-equilibrium potential, we need to introduce compensating fields which will compensate for emergence of spatial derivatives $g^{ij}(X)$ in (1). Thus, it becomes possible to construct a gauge model for the deformed state defined by the tensor $g^{ij}(X)$ in the reciprocal space. Here, we interpret a gauge model as a model in which the state is described by the pair of variables, namely the OP and its compensating field.

## 5. Scalar potential field and the first Newton's law

Let us consider a special case where a microscopic coordinate $r$ varies as follows:

$$x' = x + vt, \tag{24}$$

which corresponds to a moving reference frame with the speed $v$. Such transformation of the basis of ordinary space will be in agreement with the corresponding transformation of the basis of reciprocal space. In this case, we may assume that the transformation of the OP with some $k$

$$\hat{\vec{a}} \eta^l = e^{i(\vec{k}^l \vec{a})} \eta^l, \tag{25}$$

under the action of the temporal translation operator, will be non-trivial and will correspond to (24):

$$\hat{\tau} \eta^l = e^{i(\vec{k}^l \vec{v})\tau} \eta^l \tag{26}$$

I.e., transformation (24) results in the rotation of the basis of reciprocal space which has a projection onto the temporal axis, and this can be identified in the transformational properties of the OP (26). If we postulate such transformation of the OP for the case (24), then, given accelerated motion, a potential scalar field will inevitably emerge in such system. Let us consider a special case where $k = const$ and $v = v(T)$; here $v$ acts as the local parameter of representation. Then we will realize the necessity to extend the temporal derivative of the OP. To do that, it will be sufficient to introduce a compensating scalar field, $\phi$, which is transformed as follows:

$$\hat{\tau}(\gamma\phi) = \gamma\phi + \frac{\partial(\vec{k}\vec{v})}{\partial T} = \gamma\phi + \vec{k}\frac{\partial \vec{v}}{\partial T}\tau$$

(27)

Here, observed values are invariant combinations of the gradient of the field $\phi$ since in this case $v$ does not depend on the coordinates. The equations of state obtained from the variation of the non-equilibrium potential with respect to $\phi$, take the form similar to mathematical equations for a potential field with a charge. Here, the role of the charge is played by the density of state in the form of the first integral, for transformations from the group of temporal translations. In the present model, we may introduce the notion of inertial frame of reference which is in agreement with the first Newton's law. In fact, if $v$ does not depend on time, we do not need to introduce a compensating scalar field in the temporal derivative. Hence, the constructed local Landau theory responses to the accelerated movement of the frame of reference and requires introduction of corresponding compensating fields in the model. In the general case, the scalar $(kv)$ in Eq. (27) plays the role of frequency (26) and correlates spatial properties of the OP with $k \neq 0$ with temporal ones. Here lies the principal difference between the present model, on the one hand, and the Ginzburg-Landau model and electrodynamics as a whole, on the other hand. Refs. [1, 2, 3] considered eigenfrequency $\omega$ of the IR (10) for the temporal translation operator which was not linked with spatial transformations (24), since time is absolute and does not depend on spatial coordinates.

## 6. Conservation laws in the local Landau theory

Let us focus now on the difference between the models using either field functions [9] or local OP [2] as a fundamental variable.

As it is known, in the field theory, a multicomponent field function, $\varphi_l(X_i)$, instead of infinite number of degrees of freedom in the form of coordinates of particles, is introduced, which depends on spatial coordinates. Here we assume that under the action of the translation operator in the space $\{X_i\}$,

$$\hat{\vec{A}}\varphi_l(X_i) = \varphi_l^{'}(X_i^{'}) = \varphi_l^{'}(X_i + A_i) = \varphi_l(X_i), \qquad (28)$$

the function does not change [9]. I.e., given the transformation of the argument $X_i^{'} = X_i + A_i$, the function itself transforms in such a way that the change in the argument vanishes (28). The law of conservation of generalized momentum is associated with this transformation which leaves wave function invariant under translations.

**(Footnote:** here we denote a momentum associated with homogeneity of macrospace $\{X_i\}$ a generalized momentum, to avoid confusing it with a momentum, which is induced by a unit translation in a point of $X$ in the present model ).

Actually, we are speaking about trivial unitary representation of the subgroup of translations in the space $\{X_i\}$. This requirement is conditioned by finite-dimensionality of representations used in the field theory [9]. In the field theory [9], the scalar (vector or tensor) function is the one which transforms as an irreducible representation (IR) with $k=0$ with respect to the translation operator.

In case of a local OP, its transformation properties, as a result of action of the unit translation operator, are well-defined [2]:

$$\hat{\vec{a}}\eta_l(\vec{X}) = e^{i\vec{k}^l(\vec{X})\vec{a}}\eta_l(\vec{X}). \qquad (29)$$

In the Landau theory, the local potential must be invariant with respect to the shift by the unit cell in the physical point characterized by the macrocoordinate $X$: $\hat{\vec{a}}\,\Phi(X) = \Phi(X)$,. Here we assume

$$\hat{\vec{a}}\,(X) = X, \qquad (30)$$

where $X$ enumerates the macroscopically small volume, which we assume to be a physical point, and its coordinate does not depend on the shift by the unit period $a$ [2]. In fact, (30) is mathematical expression of the macroscopic approximation for the principle of local homogeneity. The equation (30) means fibering of space in the mathematical model, because the operator $\hat{\vec{a}}$ does not act in the space $\{X_i\}$. Here we interpret fibering not as a common mathematical term, but rather as splitting of space into subspaces in each point of $X$ where operator $\hat{\vec{a}}$ is acting. Such splitting results in mathematical fibering in the gauge models [1, 2]. Then, as it was already shown in [2], the local Landau potential is invariant with respect to transformations with three continuous parameters analogous to the 3-parameter Abelian gauge group (9), to each of them the observed first integral of the equations of state according to (22) being in correspondence.

Thus, invariance of potential with respect to translation in classical mechanics and related momentum conservation law are transformed into invariance of the inhomogeneous Landau potential with respect to unit translations (29) and related equations of continuity in the form of the second Newton's law (22).

By introducing, in order to define the state, the local OP which includes infinite number of degrees of freedom,, we should be able to consider the situation in such a way as if we shifted from conventional coordinates of particles, which were actually the degrees of freedom, to the generalized ones, $\eta_l(X_i)$, where $X_i$ now characterize macroscopic inhomogeneity. In the present model, the generalized translational degree of freedom is transformed into a phase factor of the OP that can be seen from (29). This transformation (29) is bound by the law of conservation of momentum, as shown above. The OP with trivial transformational properties under the action of the translation operators, i.e. with $k=0$ and $w=0$, does not allow minimal interactions with the compensating fields. Interactions occur upon changes in the transformational properties of the OP with the macrocoordinate, when $k=k(X)$ or $\omega=\omega(X)$, and are written down in a symmetry-dependent way in terms of the extended derivatives [1, 2].

The local gauge group of transformations for complex fields was introduced to describe interaction of the wave function with the electromagnetic potential in the field theory [9]. Phase uncertainty of the wave function allowed introducing internal symmetry of the potential, while the requirement of invariance with respect to the local group of phase transformations resulted in the gauge model of potential. However, the charge conservation law cannot be associated just with the form for representation of wave function in the complex form [9]. To the contrary, since all non-trivial irreducible representations of the subgroup of temporal translations are 2-dimensional and can be presented in the complex form, then their local representations $\omega=\omega(X)$ strongly require vector compensating fields changing their sign upon inversion of time [1,2] to be additionally introduced in the model. Then, the equations of continuity, obtained from the equations of state, contain observed values which are still retained if we integrate in macrovolume, and this results in the law of conservation of electric charge [9].

As mentioned above, in the field theory [9], the field functions are usually considered to be transformed as finite-dimensional IR, i.e. with $k=0$. For $k \neq 0$, in the general case, we deal with the infinite-dimensional IR. A similar example of infinite-dimensional representation $\eta_k$ takes place in the model considered by de Gennes in Ref. [4], where the vector $k$ of the smectic OP was deviating from the main optical axis. For $n=n(X)$ and $d=const$ the vector star $\{k\}$ represents two cones. In this case, as shown in [16], we can shift to the factorized potential which depends on two complex conjugate components of the OP, since the normal vector is defined in each point of the deformed SmA. Also it was proven in [16] that the solution for the system of equations of state exists if all components of the OP, except two complex conjugate ones, are zero. The above example with a smectic gives a clue how to deal with infinite-dimensional IR with $k \neq 0$.

Validity of the expression (30) for the considered model with local translational symmetry also stems from the E.Noether's theorem. Indeed, the equations of continuity (22) obtained from the system of equations of state (20)-(21) following the E.Noether's theorem, are connected with the transformations of functions and coordinates in the form of (29), (30) [9]. If we assume that the coordinates transform as $\hat{\tilde{a}}(X)=X+a$, then, according to the same E.Noether's theorem, we will not obtain equations of continuity (22) in the form of the second Newton's law, which exist in the present model and directly result from the equations of state (20) and (21). The expression (30) means that the unit translation $a$ is not a translation in the space $\{X\}$ in which the variational theory was formulated.

## 7. Principle of equivalence

It was shown in Section 5 how one can obtain the Newton's potential field by introducing a non-inertial frame of reference. Probably, the constructed phenomenological model of inhomogeneous states in the Landau theory of phase transitions relates to the fundamental interactions in the field theory. Newton's equations characterizing the measure of inertness of a substance result from the equations of state (20, 21) for the interacting fields: the OP and its compensating field. This interaction is characterized by the corresponding coefficient in the extended derivative before the compensating field, the phenomenological charge $\gamma(2, 19)$. From Eqs. (16, 18, 19, 20) we obtain the explicit expression for momentum in the following form:

$$p_i = -i\gamma \left[ h_i^l H_0^l - h_i^l H_0^{l*} \right]. \tag{31}$$

Here the momentum is the product of the phenomenological charge by the first integral, which corresponds to the continuous group of transformations of the potential, induced by the unit translations. The expression for momentum in the form (31) is in full agreement with the principle of equivalence of masses, if $\gamma$ corresponds to the gravity charge in (27). In the present model, momentum was defined as the charge (the first integral) and connected to gauge interaction (18). The same dependence exists in electrodynamics between unit charge in the extended derivative and the law of conservation of charge as the first integral resulting from the equations of state. Here, the nature of a unit electrical charge (phenomenological charge) and that of the retained first integral (macroscopic charge) are not viewed as different ones [9, 17]. Essentially, the result is that the equations of continuity in the gauge model (the model with symmetry-dependent interaction) are, in fact, Newton's equations of motion (22) which characterize inertial properties of the condensate. That is why there is no division into two masses in the present model: they, by definition, coincide (31).

*Footnote:* In the Landau theory, it is hardly possible to define the inert mass as the coefficient before the quadratic term of the wave function in the Lagrangian [9], since for low-symmetry state this coefficient, generally speaking, is negative.

## Conclusions

The main point of the proposed model is the concept of local homogeneity. Macroscopic approximation (30) in mathematical sense means fibering of space into subspaces in each of its points since the operator $\hat{\bar{a}}_i$, according to (30), acts in the subspace enumerated by the coordinate of $X$ but not in the space $\{X\}$. Successful results obtained in the theory of elasticity [16], [18], and successful reproduction of results of electrodynamics prove the correct description of physical phenomena which are based on the principle of local homogeneity [2]. For example, in [2, 18] the equations of state, with curl symmetry, were studied. In the present paper it is shown that the potential equations, e.g. (20), are a consequence of the constructed model. By postulating (30) we now arrive at the accurate rather than approximate results for the respective models according to the E.Noether's theorem [9]. Thus, the expression (30) means that space-time can formally be segregated into independent subspaces in each of its points in describing inhomogeneous physical states. In this case, local translational symmetry exists in each of those independent subspaces. This approach allows us to apply the gauge model where interaction is defined in terms of extended derivatives for local representations of the subgroup of translations of space-time.

The choice of the local parameter of irreducible representation and its connection with the continuous parameter of the gauge group (9, 10, and 23) is determined by the model. For irreducible local representations of the subgroup of translations, at $k=k(X, T)$, the equations of continuity correspond to the Newton's equations (22). The momentum conservation law for such model radically differs from the pulse conservation law for free field (9); it is associated with minimal interaction. Explicit dependence of momentum on phenomenological charge takes the form (31); similar dependence is known for electrical charge on unit charge in electrodynamics. Hence, in the local Landau model, momentum is proportional to phenomenological charge, by definition (20, 31), and the principle of equivalence is stemming herefrom. We obtain the Newton's potential field, when the scalar combination $(kv)$, where $v = v(T)$ (26, 27) appears as the local parameter of representation.

Note that emergence of Newton's equations, in the form of equations of continuity (22), in the local Landau theory also indicates the translational nature of the gauge group [2], since the momentum conservation law stemming from the translational invariance of nonequilibrium potential, is associated with Newton's equations.

The principle of local homogeneity does not contradict to the principles of quantum mechanics, but instead, is in agreement therewith. Since, at small distances, fields of matter have wave properties, the density of state in a small volume transformed to IR of the subgroup of translations. Indeed, the uncertainty of phase of density of probability distribution allows suggesting that the condensate has certain transformational properties in transformations from the subgroup of translations. On the other hand, the wave function $\exp(ik_j r_j - i\omega t)$ [9] is, by definition, the eigenfunction of the translation operator. Thus, the principle of local homogeneity is in full agreement with the wave-corpuscle theory which says that the density of condensate, in sufficiently small volume, has wave properties and is described by the wave function which is transformed to IR of the subgroup of translations. Localization of parameters of the wave function necessarily results in its interaction with the compensating field. The way of describing inhomogeneous states in the Landau theory corresponds to that in the quantum theory of the field, since in both cases local translational symmetry of a condensate is implied.

Hence it follows that the gauge (minimal) interaction in the field theory results from the wave nature of matter, which displays itself at very short distances. Curl invariants of compensating vector field are associated with the locality $\omega = \omega(\vec{X},t)$, consistent with the electromagnetic potential; curl tensor invariants of compensating field have the mathematical structure similar to density of dislocations in solids, and are likely to correspond to curl gravitational fields in the field theory.